\documentclass[twocolumn,showpacs,preprintnumbers,amsmath,amssymb]{revtex4}

\pdfoutput=1

\usepackage{graphicx}
\usepackage{dcolumn}
\usepackage{bm}

\begin{document}

\newcommand{\BFA}{BaFe$_2$As$_2$}
\newcommand{\KFA}{KFe$_2$As$_2$}
\newcommand{\BKFA}{(Ba$_{1-x}$K$_x$)Fe$_2$As$_2$}
\newcommand{\BKFAsf}{(Ba$_{0.6}$K$_{0.4}$)Fe$_2$As$_2$}
\newcommand{\TCS}{ThCr$_2$Si$_2$}
\newcommand{\LFAOF}{LaFeAs(O$_{1-x}$F$_x$)}

\title{Superconductivity at 38 K in the iron arsenide (Ba$_{1-x}$K$_{x}$)Fe$_2$As$_2$}

\author{Marianne Rotter, Marcus Tegel}
\author{Dirk Johrendt}
\email{johrendt@lmu.de} \affiliation{Department Chemie und Biochemie, Ludwig-Maximilians-Universit\"{a}t M\"{u}nchen, Butenandtstrasse 5-13 (Haus D), 81377 M\"{u}nchen, Germany}

\date{\today}

\begin{abstract}

The ternary iron arsenide \BFA\ becomes superconducting by hole doping, which was achieved by partial substitution of the barium site with potassium. We have discovered bulk superconductivity at $T_c$ = 38 K in \BKFA\ with $x \approx 0.4$. The parent compound \BFA\ as well as \KFA\ both crystallize in the tetragonal \TCS-type structure, which consists of (FeAs)$^{\delta-}$ iron arsenide layers separated by barium or potassium ions. \BFA\ is a poor metal and exhibits a spin density wave (SDW) anomaly at 140 K. By substituting Ba$^{2+}$ for K$^+$ ions we have introduced holes in the (FeAs)$^-$ layers, which suppress the SDW anomaly and induce superconductivity. This scenario is very similar to the recently discovered arsenide-oxide superconductors. The $T_c$ of 38 K in \BKFAsf\ is the highest critical temperature in hole doped iron arsenide superconductors so far. Therefore, we were able to expand this class of superconductors by oxygen-free compounds with the \TCS-type structure. Our results suggest, that superconductivity in these systems evolves essentially from the (FeAs)$^{\delta-}$ layers and may occur in other related compounds.

\end{abstract}


\maketitle

The recent discovery of superconductivity in pnictide-oxides with critical temperatures ($T_c$) up to 55 K has generated tremendous interest in the scientific community. After first reports on superconductivity in LaFePO \cite{Hosono-2006} and LaNiPO \cite{Hosono-2007, Tegel-2008} below 5 K, the breakthrough came with the fluoride doped arsenide \LFAOF~\cite{Hosono-2008} that exhibits $T_c$ = 26 K which increases to 43 K under pressure.\cite{Hosono-Nature} Reports on even higher $T_c$'s of up to 55 K, achieved by replacing lanthanum by rare earth ions with smaller ionic radii, followed quickly.\cite{Sm-TC55} These compounds represent the second class of high-$T_c$ materials,\cite{Angew-2008} 22 years after the discovery of the copper oxide superconductors.\cite{Bednorz-1986}

The parent compound of the new materials, LaFeAsO, has a quasi two-dimensional tetragonal structure, which consists of charged (LaO)$^{\delta+}$ layers alternating with (FeAs)$^{\delta-}$ layers (ZrCuSiAs-type).\cite{Jeitschko-1974} Several recent studies suggest that superconductivity in doped iron arsenides is unconventional and therefore non-BCS-like, \cite{Luetkens-2008,Nakai-2008,Shan-2008} but this issue is not clear at all. In contrast to the non-conducting parent compound of the copper oxides, LaFeAsO is a poor metal and exhibits Pauli-paramagnetism. The existence of a spin density wave (SDW) anomaly evolving in LaFeAsO at 135-140 K assumes a key role. \cite{Dong-SDW, Chen-SDW} The SDW is accompanied by a structural phase transition \cite{Nomura-2008} and anomalies in the specific heat, electrical resistance and magnetic susceptibility. Antiferromagnetic ordering of the magnetic moments occurs just below the structural transition temperature ($T_N$ = 134 K, 0.36 $\mu_B$/Fe).\cite{Cruz-Neutrons} By changing the electron count within the (FeAs)$^{\delta-}$ layers, the structural phase transition and antiferromagnetic ordering are suppressed and superconductivity emerges.\cite{Kitao-2008,Klauss-2008} Electron doping has been a highly successful approach in the case of LaFeAsO, either by substituting oxide for fluoride \cite{Hosono-2008} or by introducing oxide deficiencies in the LaO layer.\cite{SEFeAsO1-x} In contrast to this, the only case of superconductivity by hole doping is (La$_{1-x}$Sr$_x$)FeAsO ($T_c$ = 25 K) so far.\cite{LaSrFeAsx-25K}

\begin{figure}[h]
\includegraphics[height=50mm]{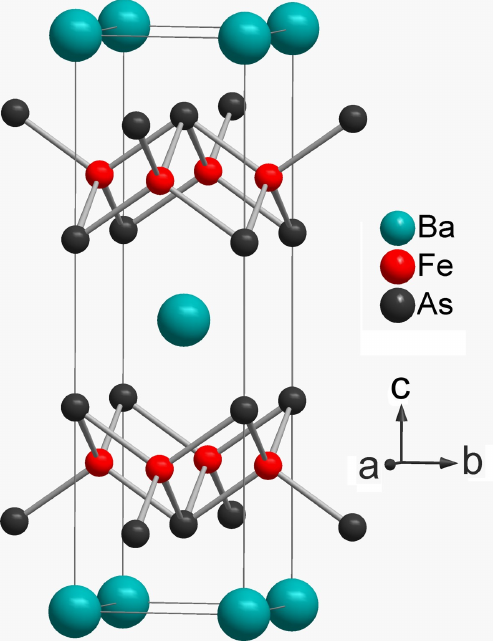}
\caption{\label{fig:Struktur}(Color online) Crystal structure of \BFA\ (\TCS-type structure, space group $I4/mmm$).}
\end{figure}

The pairing mechanism in iron arsenides is currently in dispute. But even in these early days it becomes evident, that superconductivity in LaFeAsO emerges from specific structural and electronic conditions in the (FeAs)$^{\delta-}$ layer. However, if only the iron arsenide layer is essential, also other structure types could serve as parent compounds. We reported recently, that \BFA\ with the well-known \TCS-type structure is an excellent candidate. \cite{Rotter-2008} The crystal structure of \BFA\ is shown in Figure~\ref{fig:Struktur}. This ternary arsenide contains FeAs layers identical to LaFeAsO, moreover with the same charge \cite{note_layer_charge}, and exhibits a SDW anomaly likewise (\textit{vide infra}). In this letter we report on superconductivity in \BFA\ induced by hole doping, which was achieved by partial substitution of the barium by potassium ions.

In the very large family of \TCS-type compounds, superconductivity occurred at temperatures below 5 K, as e.g. in LaIr$_2$Ge$_2$, LaRu$_2$P$_2$, YIr$_{2-x}$Si$_{2+x}$ and BaNi$_2$P$_2$, \cite{Shelton-1984,Venturini-1985,Jeitschko-1987,Hagenmuller-1985,BaNi2P2-2008} although closely related rare earth borocarbides are known for higher $T_c$'s up to 26 K in YPd$_2$B$_2$C.\cite{Cava-1994, Borocarbides} We have suggested earlier, that superconductivity in \TCS-type compounds may arise under certain electronic conditions. \cite{Johrendt-1997} Recently we found that \BFA\ and LaFeAsO exhibit strikingly similar properties. \cite{Rotter-2008} \BFA\ is a poor Pauli-paramagnetic metal that undergoes a structural and magnetic phase transition at 140 K, accompanied by strong anomalies in the specific heat, electrical resistance and magnetic susceptibility. In the course of this phase transition, the space group symmetry changes from tetragonal ($I4/mmm$) to orthorhombic ($Fmmm$).

Based on these findings, we expected superconductivity in doped \BFA. First attempts to realize electron doping by lanthanum substitution were unsuccessful, because the required doping level could not be achieved. We then decided to try hole doping by substituting the Ba$^{2+}$ cations for K$^+$ with a similar ionic radius. In order to achieve doping levels of 0.15-0.2 electrons per (FeAs) unit, we had to substitute 30-40\% of the barium ions for potassium ions. This seemed possible, because isostructural \KFA\ had been known to exist.\cite{Schuster-1981} We succeeded in preparing \BKFA\ ($x$ = 0.3 and 0.4) by heating stoichiometric mixtures of the elements (all purities $>$ 99.9\%) in alumina crucibles, welded in silica tubes under an atmosphere of purified argon. \cite{Nagorsen-1980,Schuster-1981} The samples were heated slowly (50 K/h) to 873 K, kept at this temperature for 15 h and cooled to room temperature by switching off the furnace. After homogenization in an argon-filled glove-box, the products were annealed at 925 K for 15 h, again homogenized, cold  pressed into pellets and sintered at 1023 K for 12 h. The material is black and stable in air for weeks. \BFA\ and \KFA\ were synthesized by the same method.

\begin{figure}[h]
\includegraphics[width=85mm]{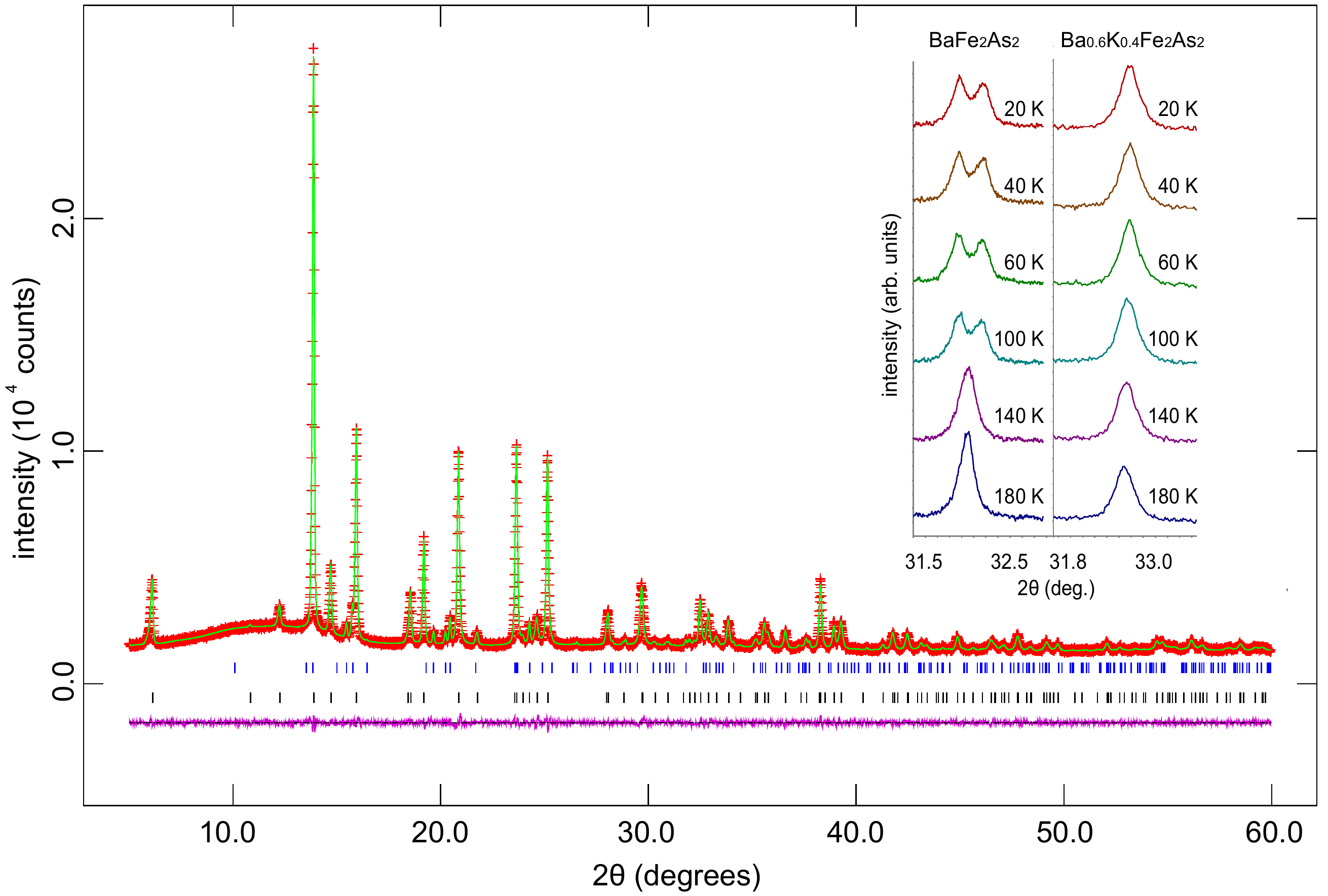}
\caption{\label{fig:Rietveld} (Color online) X-ray powder pattern (+) and Rietveld fit ($-$) of \BKFAsf\ at 297 K. Reflection markers are blue for FeAs and black for \BKFAsf. The insert shows the temperature dependency of the (110)-reflections of \BFA\ and \BKFAsf.}
\end{figure}

Phase purity was checked by X-ray powder diffraction with Cu-$K_{\alpha_{1}}$ or Mo-$K_{\alpha_{1}}$ radiation. Rietveld refinements of the data were performed with the GSAS package \cite{GSAS,Finger-Cox-Jephcoat}. The pattern of \BFA\ could be completely fitted with a single phase. In the samples of \KFA\ and \BKFA, we detected FeAs (Westerveldite \cite{Selte-FeAs}) as impurity phase, which was included in the refinement and quantified to $6\pm1\%$. The substitution of 40\% barium for potassium is clearly proved by the refinement of the site occupation parameters in the Rietveld fit of \BKFAsf\ (Figure~\ref{fig:Rietveld}). A summary of the crystallographic data is compiled in Table~\ref{tab:Crystallographic}.

As mentioned above, a crucial aspect of the LaFeAsO superconductors is the suppression of the SDW anomaly by doping. Therefore, we have also measured the X-ray powder pattern of \BKFAsf\ at 20 K. No broadening or splitting of the diffraction peaks as found in \BFA\ below 140 K were detected. The insert in  Figure~\ref{fig:Rietveld} shows the temperature dependency of the (110)-reflections of \BFA\ and \BKFAsf\ for comparison. We successfully refined the pattern of \BKFAsf\ measured at 20 K by using the parameters of the undistorted tetragonal structure (space group $I4/mmm$). Table~\ref{tab:Crystallographic} shows the almost identical crystallographic data of \BKFAsf\ at 297 K and 20 K, respectively. Thus it is evident that the K-doping has suppressed the structural transition of \BFA.

\begin{table}[h]
\caption{\label{tab:Crystallographic} Crystallographic data of \BKFAsf}
\begin{ruledtabular}
\begin{tabular}{lll}
Temperature (K) & 297    & 20 \\
 Space group & $I4/mmm$           & $I4/mmm$ \\
 \textit{a} (pm) & 391.70(1)        & 390.90(1) \\
 \textit{c} (pm) & 1329.68(1)       & 1321.22(4) \\
 \textit{V} (nm$^{3}$) & 0.20401(1) & 0.20189(1) \\
 \textit{Z} & 2                   & 2 \\
 data points & 5499               & 8790 \\
 reflections (total) & 405        & 127 \\
 \textit {d} range & $0.639 - 6.648$ & $0.971 - 6.606$ \\
 R$_\text{P}$, \textit{w}R$_\text{P}$ & 0.0202, 0.0258 & 0.0214, 0.0283\\
 R$(F2)$, $\chi2$ & 0.026, 1.347 & 0.093, 1.816\\
Atomic parameters: \\
K,Ba & 2$a$ (0,0,0)                       &  2$a$ (0,0,0)\\
    & $U_{iso} = 130(8)$                   & $U_{iso} = 89(8)$            \\
Fe & 4$d$ ($\frac{1}{2},0,\frac{1}{4}$)   &  4$d$ ($\frac{1}{2},0,\frac{1}{4}$)\\
    & $U_{iso} = 47(4)$                    & $U_{iso} = 84(7)$\\
As & 4$e$ (0,0,$z$)                       &  4$e$ (0,0,$z$)  \\
    & $z$ = 0.3538(1)                      &  $z$ = 0.3538(1) \\
    & $U_{iso} = 70(3)$                    & $U_{iso} = 76(7)$\\
K : Ba ratio & 42(1) : 58(1)              & 38(1) : 62(1) \\
Bond lengths (pm):\\
Ba--As  &  338.4(1)$\times$8          & 337.2(1)$\times$8\\
Fe--As  &  239.6(1)$\times$4          & 238.8(1)$\times$4\\
Fe--Fe  &  277.0(1)$\times$4          & 276.4(1)$\times$4\\
Bond angles (deg):\\
As--Fe--As &  109.7(1)$\times$2       & 109.9(1)$\times$2\\
           &  109.4(1)$\times$4       & 109.3(1)$\times$4\\
\end{tabular}
\end{ruledtabular}
\end{table}

\begin{figure}[h]
\includegraphics[width=90mm]{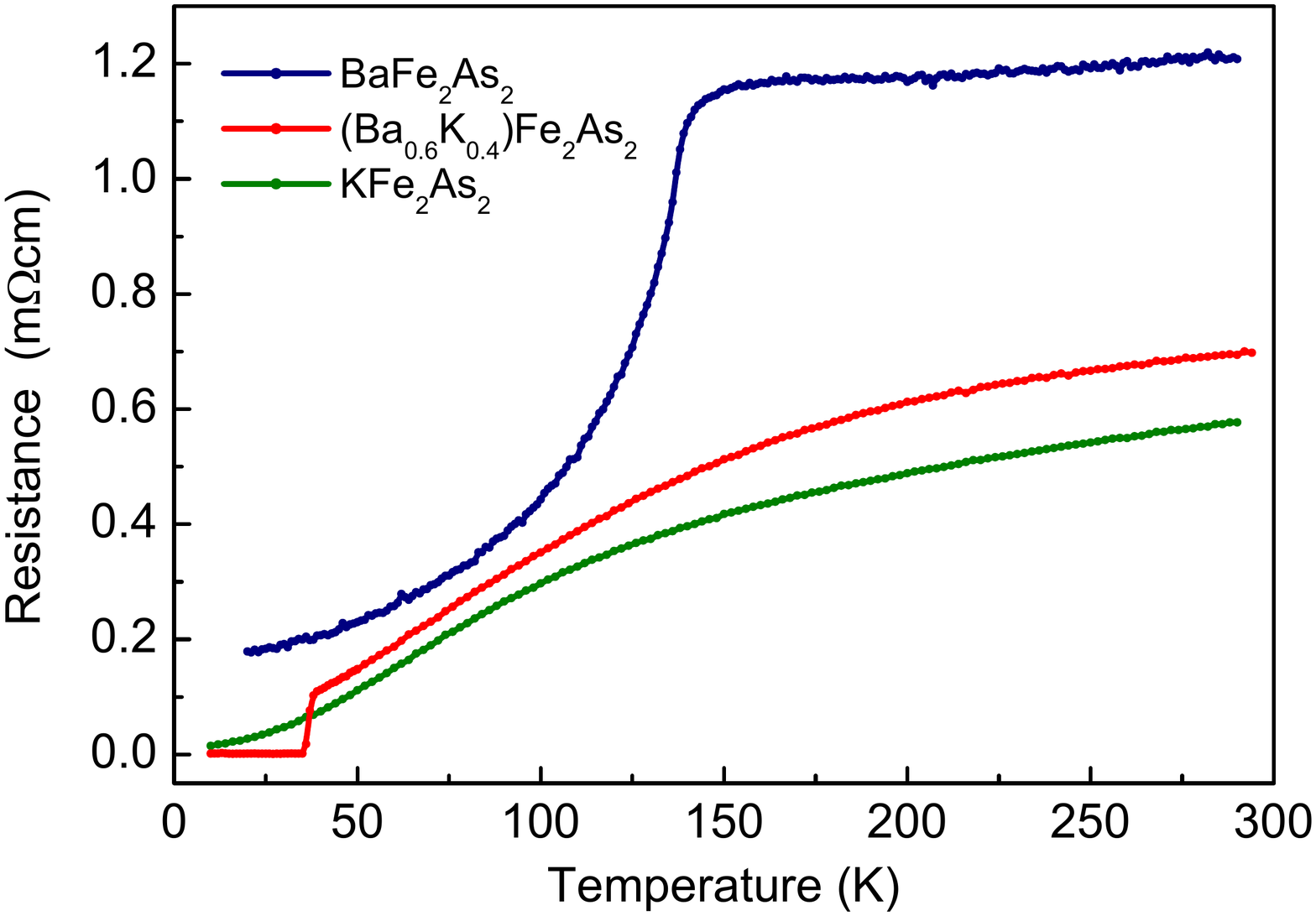}
\caption{\label{fig:RHO}(Color online) Electrical resistance of \BFA\ (blue), \KFA\ (green) and \BKFAsf\ (red).}
\end{figure}

\begin{figure}[h]
\includegraphics[width=90mm]{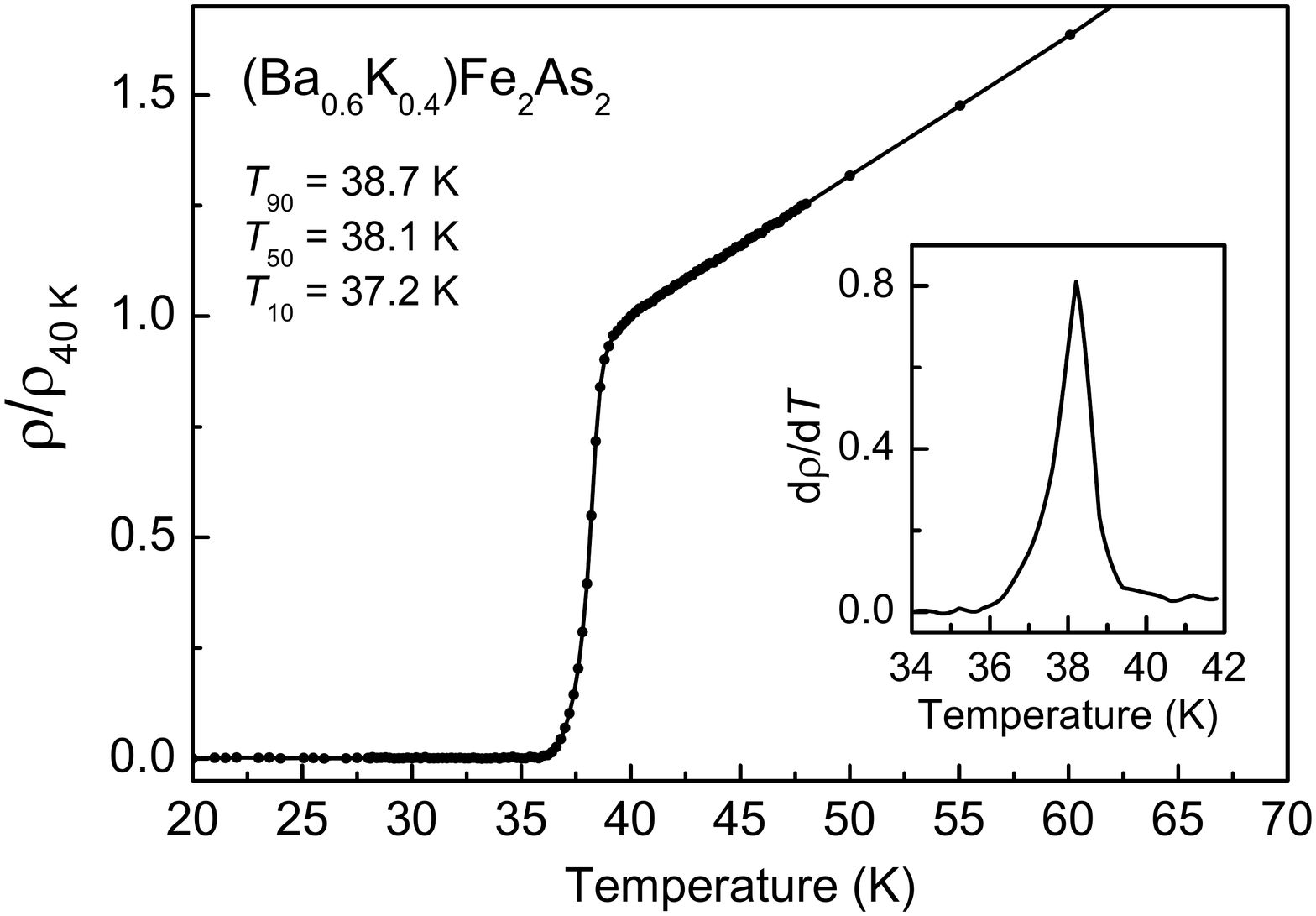}
\caption{\label{fig:RHO2}Resistivity transition of \BKFAsf.}
\end{figure}

That followed we have measured the electrical resistance of \BKFA\ ($x$ = 0, 0.4 and 1.0) through a four-probe method. As depicted in Figure~\ref{fig:RHO}, \BFA\ has the highest resistance and shows a decrease at 140 K, which is linked to the SDW anomaly.\cite{Rotter-2008} In contrast to this, the resistance of \KFA\ is considerably smaller and decreases smoothly with temperature, as it is typical for a normal metal. The resistance of K-doped \BKFAsf\ is similar to \KFA\ and does not show any sign of an anomaly at about 140 K in agreement with our structural data. But the resistance drops abruptly to zero at $\approx 38$ K, which clearly indicates superconductivity. Figure~\ref{fig:RHO2} shows details of the transition. By using the 90/10 criterion, we find the midpoint of the resistive transition at 38.1 K and a transition width of 1.5 K. The first deviation from the extrapolated resistance is at $\approx$ 39 K and zero resistance is achieved at 37.2 K. Consequently, we have discovered superconductivity analogue to the LaFeAsO materials, but in a oxygen-free compound with \TCS\ structure. The $T_c$ of 38 K is the highest critical temperature observed in hole doped iron arsenide superconductors so far (25 K in (La$_{1-x}$Sr$_x$)FeAsO).\cite{LaSrFeAsx-25K}

\begin{figure}[h]
\includegraphics[width=90mm]{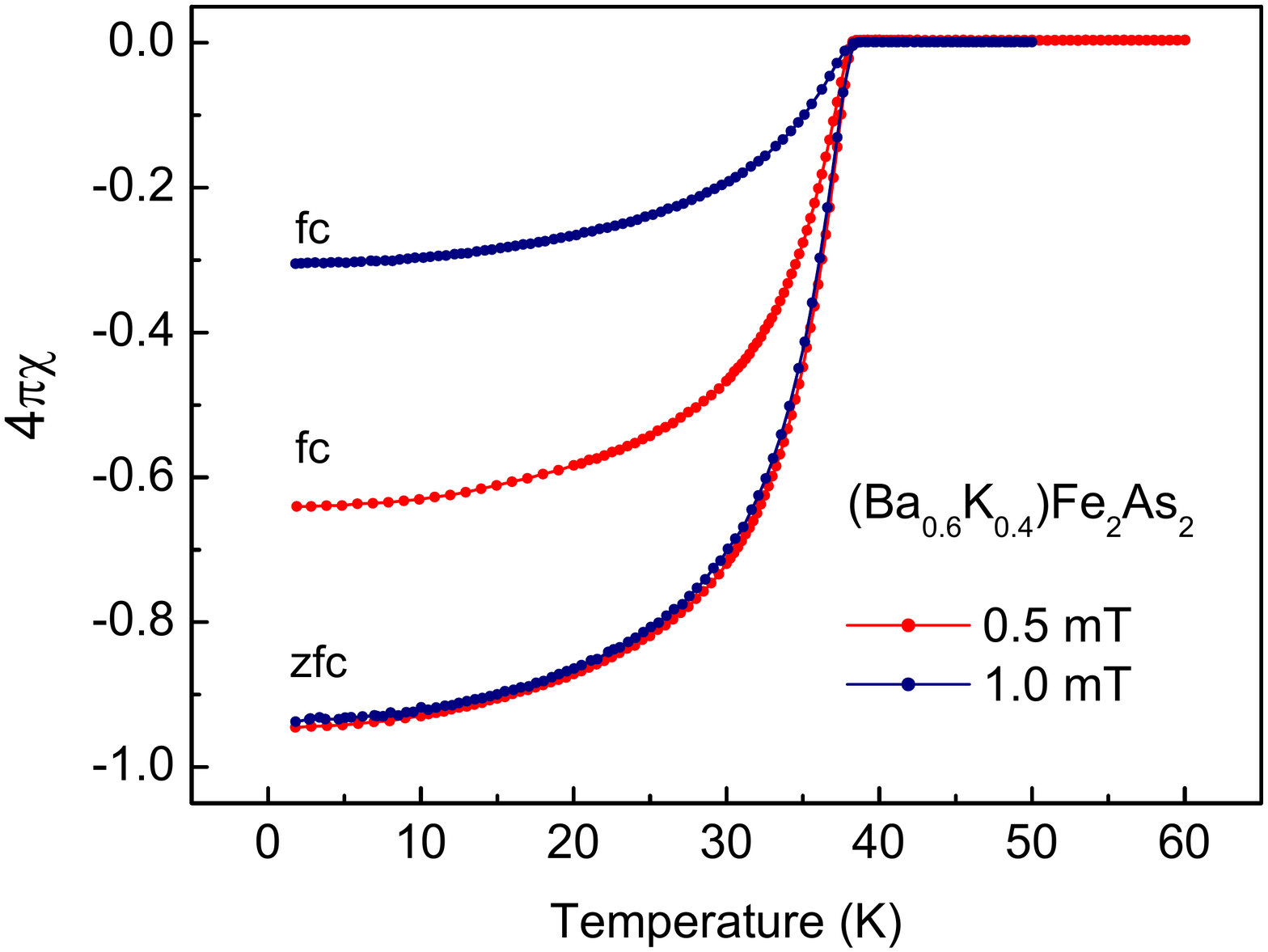}
\caption{\label{fig:CHI}(Color online) Magnetic susceptibility of \BKFAsf\ at 0.5 mT (red) and 1 mT (blue).}
\end{figure}

In order to confirm superconductivity, we have measured the magnetic susceptibility of finely ground powder of \BKFAsf\ using a SQUID magnetometer (MPMS-XL5, Quantum Design Inc.) Zero-field cooled (shielding) and field cooled (Meissner) cycles measured at 1 mT and 0.5 mT are shown in Figure~\ref{fig:CHI}. The sample becomes diamagnetic at 38.3 K and shows 10\% of the maximum shielding at 37.2 K. The zero field cooled (zfc) branches of the susceptibilities measured at 1 and 0.5 mT are almost identical and amount to $-0.94$ at 1.8 K, which is close to ideal diamagnetism (4$\pi\chi = -1$). The Meissner effect (fc) depends on the applied field and the measured susceptibilities at 1.8 K are $-0.64$ at 0.5 mT and $-0.3$ at 1 mT. These values of the shielding- and Meissner fractions should be considered as estimates due to uncertainties regarding the density of the compacted powder and demagnetization effects. However, the susceptibility data unambiguously prove bulk superconductivity of the \BKFAsf\ sample.

In summary, we have discovered the first member of a new family of iron arsenide superconductors. \BKFAsf\ with the \TCS-type structure is a bulk superconductor with $T_c$ = 38 K. The structural and electronic properties of the parent compound \BFA~are closely related to LaFeAsO. We have induced superconductivity by hole doping and have found a significantly higher $T_c$ in comparison with hole doped LaFeAsO. In contrast to previously stated opinions, our results suggest that hole doping is definitely a possible pathway to induce high-$T_c$ superconductivity, at least in the oxygen-free compounds. Further optimization may lead to even higher $T_c$'s in \TCS-type compounds.

\begin{acknowledgments}
We thank Prof. Thomas F\"{a}ssler for support with magnetic measurements. This work was financially supported by the German Research Foundation [Deutsche Forschungsgemeinschaft (DFG)].
\end{acknowledgments}

\bibliographystyle{apsrev}


\end{document}